\documentclass[showpacs,preprintnumbers,prc,twocolumn]{revtex4}
\usepackage{graphicx}
\usepackage{amsmath,amssymb,amsfonts}
\usepackage{array}
\usepackage{url}
\usepackage{hyperref}
\usepackage{multirow}
\usepackage{float}
\usepackage{lineno}
\usepackage{xspace}
\usepackage[usenames,dvipsnames]{color}

\newcommand{\bef}{\begin{figure}}
\newcommand{\eef}{\end{figure}}
\newcommand{\bc}{\begin{center}}
\newcommand{\ec}{\end{center}}

\newcommand{\be}{\begin{equation}}
\newcommand{\ee}{\end{equation}}
\newcommand{\bea}{\begin{eqnarray}}
\newcommand{\eea}{\end{eqnarray}}

\def\ba{\begin{eqnarray}}
\def\ea{\end{eqnarray}}
\begin{document}

\title{Thermodynamic and Transport Properties of Matter Formed in $pp$, $p$-Pb, Xe-Xe and Pb-Pb Collisions at the Large 
Hadron Collider using Color String Percolation Model}
\author{Dushmanta Sahu}
\author{Raghunath Sahoo\footnote{Corresponding Author Email: Raghunath.Sahoo@cern.ch, Presently CERN Scientific Associate at CERN, Geneva, Switzerland}}
\affiliation{Department of Physics, Indian Institute of Technology Indore, Simrol, Indore 453552, India}

\begin{abstract}
To have a better understanding of the matter formed in ultra-relativistic collisions, using the Color String Percolation Model (CSPM), we have estimated initial energy density, mean free path, squared speed of sound, shear viscosity to entropy density ratio ($\eta/s$), bulk viscosity to entropy density ratio and bulk modulus for $pp$, $p$-Pb, Xe-Xe and Pb-Pb collisions at the Large Hadron Collider (LHC). These observables are studied as a function of final state charged particle multiplicity density normalized by nuclear overlap area and initial state percolation temperature. The results from this work are compared with the results from well known QCD models and with the transport properties of matter found in day-to-day life. These observables allow us to conclude an array of important findings about the matter formed in high-energy collisions at the LHC. 
For the high-multiplicity events, the estimated $\eta/s$ are comparable with that is observed for heavy-ion collisions, which is an indication of  a strongly coupled Quark Gluon Plasma. This makes LHC high-multiplicity $pp$ collision events as probable candidates for the search of QGP droplets. 
 \pacs{}
\end{abstract}
\date{\today}
\maketitle

\section{Introduction}
\label{intro}
With the development of high-energy particle colliders at CERN and BNL, the interest and intrigue about the matter formed in ultra-relativistic collisions have been the driving forces of recent research in high-energy nuclear physics. Since the last few decades, scientists have made various discoveries on the possible formation of a new state of matter called Quark-Gluon Plasma (QGP). QGP is a de-confined state of matter with quarks and gluons as its degrees of freedom. It is a result of first order/cross-over phase transition from normal baryonic matter at extremely high temperature and/or high density. The earlier mainstream belief was that QGP could be formed only in heavy-ion collisions. But recent studies have shown a glimpse about the possible formation of QGP droplets even in high-multiplicity $pp$ collisions~\cite{nature, Sahoo:2019ifs, Velicanu:2011zz}. To confirm, one needs to gather substantial evidence of a QGP-like medium formation in high-multiplicity pp collisions. Thus, it is important to learn about the thermodynamic and transport properties of matter formed in such collisions and compare the estimations from other collision species like $p$-Pb, Xe-Xe and Pb-Pb collisions at the Large Hadron Collider (LHC). For the completeness sake and as a systematic study, we also estimate the thermodynamic and transport properties for low-multiplicity $pp$ collisions, however, the physical interpretation of the low-multiplicity system has to be done with caution. 
The charged particle multiplicity density normalized with nuclear overlap area ($S_{\rm N}$) and proper time ($\tau_{0}$) is a better proxy of the density of the system. In the current study, we assume $\tau_{0} \sim$ 1 fm/c and estimate the dependency of the charged particle multiplicity density ($\langle dN_{ch}/d\eta \rangle$) normalized by nuclear overlap area on  thermodynamic and transport properties such as energy density ($\epsilon$), mean free path ($\lambda$), shear viscosity to entropy density ratio ($\eta/s$), speed of sound ($c_{s}$), bulk viscosity to entropy density ratio ($\zeta/s$) and Bulk Modulus (B) using a Color String Percolation Model (CSPM). $\langle dN_{ch}/d\eta \rangle$ is the charged particle multiplicity per unit pseudorapidity at the midrapidity, where the average signifies the event average for a given class of events as defined in Refs. \cite{ALICE:2019etb,Acharya:2018orn,Abelev:2013vea,Acharya:2019yoi,Acharya:2019rys,Acharya:2018hhy,Acharya:2018egz}. 
It would be interesting to compare these results with the estimations from lattice QCD~\cite{Karsch:2007jc}, excluded volume hadron resonance gas model~\cite{Tiwari:2011km} and the ideal gas Stefan-Boltzmann limit.

The initial energy density produced in high-energy collisions is an important parameter, which not only gives a first hand information
about possible deconfinement phase through a direct comparison with the critical energy density by lattice QCD estimations, but
also drives the subsequent spacetime evolution of the fireball, while controlling the system dynamics and particle production.
Mean free path ($\lambda$) is an important thermodynamical quantity which is defined as the average distance travelled by a particle between two successive collisions. The magnitude of mean free path depends on the characteristics of a system. As the density of the system increases, its mean free path decreases. This gives us an idea about how the system behaves at extreme conditions of temperatures. The shear viscosity to entropy density ratio helps in determining a threshold of charged particle multiplicity density after which one can measure the fluidity of a system \cite{Lacey:2006bc}. The elliptic flow measurements from heavy-ion collisions at RHIC \cite{STAR-NPA} have found that the system formed in heavy-ion collisions at the LHC gives $\eta /s$ value closer to the KSS (Kovtun-Son-Starinets) bound, which might suggest that QGP is almost a perfect fluid \cite{Romatschke:2007mq,Hirano:2005wx}. Similarly, a change in behavior of bulk viscosity to entropy density ratio ($\zeta/s$) is expected near QCD critical temperature ($T_c$),  where  conformal  symmetry  breaking  might be significant as predicted by several effective models~\cite{Dobado:2012zf,Sasaki:2008fg,Shushpanov:1998ce}. Speed of sound uncovers the strength of interactions of the constituents of a medium. For a massless ideal gas, the squared speed of sound ($c_{\rm s}^{2}$) is expected to be 1/3 in
the picture of Landau hydrodynamics describing multihadron production dynamics \cite{Landau}, whereas for an hadron gas the value is expected to be 1/5 \cite{Mohanty:2003va}. A comparison with the results from this work to the massless ideal gas and hadron gas would give a hint about the system dynamics. The bulk modulus (B) is the inverse of isothermal compressibility, which gives an indication of resistance of the system against any compression. Similar to $\eta /s$, the bulk modulus can help in measuring the fluidity of a system.

To study various thermodynamic and transport properties of the matter formed in ultra-relativistic collisions at the LHC, we have used the framework of well established Color String Percolation Model (CSPM)~\cite{Braun:2015eoa}. Similar to Color Glass Condensate (CGC) and hydrodynamical descriptions of QCD \cite{McLerran:1993ni}, the CSPM gives a reasonable description of several thermodynamic and transport properties. CSPM is a QCD inspired model, however, it is not directly derived from QCD. It is an alternative approach to the CGC, which is described in terms of the percolation of color strings. In CSPM, it is assumed that color strings are stretched between the target and the projectile. These color strings decay into new color strings through the production 
of color neutral quark-antiquark pairs. 
These strings have a transverse area $S_{\rm 1}$ and longitudinal length $\sim$ 1 fm. They seem like color flux tubes stretched between colliding partons. The number of color strings grows with growing energy and the increase in the number of colliding partons. It results in color tubes overlapping with each other in the transverse space. This increases the string density and string tension of the composite strings. The net color in the overlapping string area is the vector sum of randomly oriented colored strings.
After a certain critical percolation density ($\xi_{\rm c} \sim$ 1.2),  when more than half of the surface area is covered by the colored strings, a macroscopic cluster appears which marks the percolation phase transition ~\cite{Braun:2015eoa,Satz:2000bn}. 
In the process, Schwinger string breaking mechanism produces color neutral pairs of quark-antiquark in time $\sim$ 1 fm/c from the QCD vacuum, which subsequently hadronize to form the final state particles \cite{Schwinger,Wong0,Wong}.

In this work, we have estimated thermodynamic and transport properties for $pp$, $p$-Pb, Xe-Xe and Pb-Pb collisions using the CSPM approach. These observables are studied as a function of final state charged particle multiplicity density (event classifier used in LHC) normalized by nuclear overlap area and initial state percolation temperature. The results from this work are compared to the results from well known models of QCD and also they are compared to the thermodynamic and transport properties of matter found in day-to-day life such as water. The primary aim of this work is to express different thermodynamic observables as a function of the event classifier ($\langle dN_{ch}/d\eta \rangle$) used in ALICE at the LHC  normalized to the nuclear overlap area and include the results from $pp$, $p$-Pb, Xe-Xe and Pb-Pb collisions at different center-of-mass energies for a specific thermodynamic observable. This also helps to study various observables across different collision species and collision energies as a function of final state event multiplicity. 
The scaling of final state multiplicity is supported by the universality/similarity in hadroproduction observed in hadronic and heavy-ion collisions across a broad range of collision energies \cite{EKG,Mishra:2014dta,Sarkisyan:2015gca,Sarkisyan:2016dzo,Sarkisyan-Grinbaum:2018yld}. Further the thermodynamic and transport properties are also studied as a function of initial percolation temperature normalized by the critical temperature.

The paper is organised as follows. In section \ref{formulation}, we have given a brief formulation of different thermodynamic and transport properties in the CSPM framework. In section \ref{res}, the results are discussed and they are summarized in section \ref{sum}.

\section{Formulation}
\label{formulation}

Before going to the estimation of various thermodynamic and transport properties, let us first discuss briefly about the Color String Percolation Model (CSPM). In a 2D percolation theory, the dimensionless percolation density parameter is written as,
\begin{equation}
\label{eq1}
\xi = \frac{N_{\rm s}S_{\rm 1}}{S_{\rm N}}.
\end{equation}
Here $N_{\rm s}$ is the number of overlapping strings, $S_{\rm 1}$ is the transverse area of a single string and $S_{\rm N}$ is given by the transverse area of the overlapping strings \cite{Braun:2002ed}. We have taken the values of $S_{\rm N}$ for Pb-Pb, Xe-Xe and $p$-Pb collisions from Ref. \cite{Loizides:2017ack}. For the calculation of $S_{\rm N}$ for $pp$ collisions, the radius is calculated from IP-Glasma model \cite{McLerran:2013oju,Hirsch:2018pqm}.

We evaluate $\xi$ by fitting the experimental data of $pp$ collisions at $\sqrt s$ = 200 GeV with the following function~\cite{BKnpa},
\begin{equation}
\label{eq2}
\frac{d^{2}N_{\rm ch}}{dp_{\rm T}^2} = \frac{a}{(p_{\rm 0}+p_{\rm T})^{\alpha}},
\end{equation}
where $a$ is the normalization factor. $p_{\rm 0}$ and $\alpha$ are the fitting parameters, which are given as $p_{\rm 0}$ = 1.982 GeV and $\alpha$ = 12.877. To evaluate the interaction of strings in high-energy $pp$, $p$-A and A-A collisions, we update the parameter $p_{\rm 0}$ as \cite{Braun:2002ed},
\begin{equation}
\label{eq3}
p_{0} \rightarrow p_{0} \bigg(\frac{\langle \frac{N_{s}S_{1}}{S_{N}}\rangle_{pp,pA,AA}}{\langle \frac{N_{s}S_{1}}{S_{N}}\rangle_{pp, \sqrt s = 200 GeV}}\bigg)^{1/4}.
\end{equation}
In the thermodynamic limit, where $N_{\rm s}$ and $S_{\rm N}$ $\to$ $\infty$, keeping $\xi$ fixed we get,
\begin{equation}
\label{eq4}
\bigg\langle\frac{N_{s}S_{1}}{S_{N}}\bigg\rangle = \frac{1}{F^{2}(\xi)}.
\end{equation}
Here, the average is taken over the number of cluster of colored strings. $F(\xi)$ is the color suppression factor and is related to $\xi$ by the relation
\begin{equation}
\label{eq5}
F(\xi) = \sqrt{\frac{1-e^{-\xi}}{\xi}}.
\end{equation}

Using the Eq.(2), for $pp$, $p$-Pb, Xe-Xe and Pb-Pb collision systems at LHC energies, we obtain,
\begin{equation}
\label{eq6}
\frac{d^{2}N_{ch}}{dp^{2}_{T}} = \frac{a}{(p_{0}\sqrt{F(\xi)_{pp, \sqrt s = 200 ~{\rm GeV}}/F(\xi)_{pp,pA,AA}}+p_{\rm T})^{\alpha}}.
\end{equation}

Due to low string overlap probability in low energy $pp$ collisions, we assume $\big\langle \frac{N_{s}S_{1}}{S_{N}}\big\rangle \sim 1$. We have used the above equation to fit the softer part of the $p_{\rm T}$ spectra in the range 0.12 - 1.0 GeV/c, from which we obtain the multiplicity dependent $F(\xi)$ values for different collision systems. From Schwinger mechanism of particle production \cite{pajares3}, the initial percolation temperature can be written in terms of $F(\xi)$ as,

\begin{equation}
\label{eq7}
T(\xi) = \sqrt{\frac{\langle p^{2}_{\rm T}\rangle_{1}}{2F(\xi)}}.
\end{equation}
By using $T_{c} = 167.7\pm2.8$ MeV~\cite{Becattini:2010sk} and $\xi_{c} \sim 1.2$, we get $\sqrt{\langle p^{2}_{\rm T}\rangle_{1}} = 207.2\pm3.3$~MeV. We can estimate the initial temperature for different $F({\xi})$ values by using the above values in Eq. \ref{eq7}. To estimate various thermodynamical observables, we have used the obtained temperature and $\xi$ values, whereas the $\langle dN_{\rm ch}/d\eta \rangle$ values are taken from Refs.~\cite{ALICE:2019etb,Acharya:2018orn,Abelev:2013vea,Acharya:2019yoi,Acharya:2019rys,Acharya:2018hhy,Acharya:2018egz}. In the following subsections, we briefly describe the formulation of different thermodynamic observables in CSPM approach.

\subsection{Initial Energy Density}

When the initial percolation temperature is higher than the critical temperature ($T > T_{c}$), the CSPM fluid expands according to Bjorken boost invariant 1-D hydrodynamics. Thus, the Bjorken energy density in CSPM approach is given by \cite{Bjorken:1982qr,Braun:2015eoa,Sahoo:2018dcz},
\begin{equation}
\label{eq13}
\epsilon = \frac{3}{2}\frac{dN_{\rm ch}/dy \langle m_{\rm T}\rangle}{S_{\rm N}\tau_{\rm pro}},
\end{equation}
where $S_{\rm N}$ is the transverse overlap area, $m_{\rm T} = \sqrt{m^{2}+p_{\rm T}^{2}}$ is the transverse mass, with $m$ being the mass of pion, as pion is the most abundant particle in a multiparticle production process. $\tau_{\rm pro}$ is the parton production time which is assumed to be $\sim$ 1 fm/c and $\tau_{\rm pro} = \frac{2.405\hbar}{\langle m_{\rm T}\rangle}$. Here we have assumed $dN_{\rm ch}/dy$ to be $dN_{\rm ch}/d\eta$ by assuming a Bjorken rapidity and pseudorapidity correlation. The conversion of rapidity to pseudorapidity is associated with a Jacobian. If the rest mass of the particle is very small than the particle momenta, the Jacobian becomes 1. Thus, the value of the Jacobian is smaller at LHC (1.09 for central Pb-Pb collisions at $\sqrt{s_{\rm NN}} = 2.76$ TeV \cite{CMS:2012krf}) as compared to the RHIC energies (1.25 for central Au-Au collisions at $\sqrt{s_{\rm NN}} = 200$ GeV \cite{PHENIX:2004vdg}). For  the sake of simplicity, we have assumed the Jacobian to be 1 for all the considered systems in this work, which is a fair assumption.
By using Eq \ref{eq13}, we have estimated the initial percolation energy density for various systems at different collision energies.

\subsection{Mean Free Path ($\lambda$)}
In thermodynamics, the mean free path inside a medium is defined as
\begin{equation}
\label{eq8}
\lambda = \frac{1}{n\sigma},
\end{equation}
where $n$ is the number density and $\sigma$ is the scattering cross-section of the particles. For simplicity, in this work it is assumed that the particles collide elastically with each other. By using CSPM, we obtain the mean free path using the following equation \cite{Braun:2015eoa,Sahoo:2017umy},
\begin{equation}
\label{eq9}
\lambda = \frac{L}{(1-e^{-\xi})}. 
\end{equation}
Here $L$ is the longitudinal length of the string $\sim$1 fm. One can observe from the above equation that for very large values of $\xi$, the mean free path becomes constant ( $\sim$ 1 fm). 

\subsection{Shear Viscosity to Entropy Density Ratio}

Shear viscosity to entropy density ratio is a measure of the fluidity of the system. From the ADS/CFT calculations a lower bound (KSS bound) to the shear viscosity to entropy density ratio ($\eta/s$) for any fluid has been estimated. The elliptic flow measurements from heavy-ion collisions at Relativistic heavy-ion collisions (RHIC) at BNL found that the medium formed in such collisions are closer to the KSS bound ($\simeq 1/4\pi$), which might indicate that the matter formed in such collisions behave as nearly perfect fluid  \cite{Kovtun:2004de}. At RHIC, the shear viscosity of the matter formed is a time-dependent quantity. This suggests that, in the partonic phase of such collisions, we would expect a very low value of $\eta/s$ and after hadronization, it is expected to increase. Several investigations lead to the fact that in the vicinity of a first-order/cross-over phase transition, $\eta/s$ should reach a minimum, but then again starts increasing in the deconfined phase \cite{Csernai:2006zz}. 

In the CSPM approach, the shear viscosity to entropy density ratio can be expressed as~\cite{Sahoo:2017umy}:
\begin{equation}
\label{eqa}
\eta/s \simeq \frac{T\lambda}{5} \nonumber\\
\end{equation}
We then substitute the expression of mean free path from Eq.\ref{eq9} in the above equation. The final expression for the shear viscosity to entropy density ratio is given by,

\begin{equation}
\label{eq10}
\eta/s = \frac{TL}{5(1-e^{-\xi})}.
\end{equation}

\subsection{Speed of Sound}
The speed of sound gives the information about the equation of state of the system. It describes the conversion of the change in energy density profile of the created medium into pressure gradient. In hydrodynamics, collective expansion is observed because of the pressure gradient created in the medium. From boost-invariant Bjorken 1D hydrodynamics, the squared speed of sound is given by \cite{Braun:2015eoa},

\begin{equation}
\label{eqb}
\frac{1}{T}\frac{dT}{d\tau} =  -c_{s}^{2}/\tau,
\end{equation}
where $\tau$ is the proper time. The proper time derivative of temperature can be otherwise written as,
\begin{equation}
\frac{dT}{d\tau} =  \frac{dT}{d\epsilon}\frac{d\epsilon}{d\tau}. \nonumber\\
\end{equation}
By differentiating the initial energy density with respect to proper time, we get
\begin{equation}
\frac{d\epsilon}{d\tau} =  -Ts/\tau, \nonumber\\
\end{equation}
where $s$ is the entropy density, which is given by $s = (\epsilon + P)/T$, and $P = (\epsilon - \Delta T^{4})/3$. By using the above expressions in Eq.\ref{eqb} and simplifying we get,
\begin{equation}
\frac{dT}{d\epsilon}s =  c_{s}^{2}. \nonumber\\
\end{equation}
In CSPM approach, the above expression can be simplified and written in terms of the percolation density parameter ($\xi$) as,

\begin{equation}
c_{s}^{2} = (-0.33)\bigg(\frac{\xi e^{-\xi}}{1 - e^{-\xi}} - 1\bigg) \nonumber\\
\end{equation}
\begin{equation}
\label{eq11}
+ 0.0191(\Delta/3)\bigg(\frac{\xi e^{-\xi}}{(1 - e^{-\xi})^2} - \frac{1}{1 - e^{-\xi}}\bigg),
\end{equation}
where, $\Delta = (\epsilon - 3P)/T^{4}$ is the trace anomaly.

\subsection{Bulk Viscosity to Entropy Density Ratio}

Bulk viscosity ($\zeta$) or volume viscosity is the property that characterizes flow of a fluid. The bulk viscosity is relevant only if the density of the medium is changing. It plays an important role in attenuating sound waves in fluids and can be estimated from the magnitude of the attenuation. For an almost incompressible fluid, the changes in density can be ignored. In the perfect fluid limit the energy density decreases with proper time as a consequence of longitudinal expansion. In the meantime, the viscosity opposes the system to perform useful work while expanding longitudinally. The study of bulk viscosity to entropy density ratio ($\zeta/s$) is very crucial as it can quantify the critical charged particle multiplicity density after which a change in the dynamics of the system can be observed.

From the relaxation time approximation, the bulk viscosity of a system is given as \cite{Weinberg:1971,Dusling:2011fd},
\begin{equation} \nonumber\\
\zeta = 15\eta\bigg(\frac{1}{3}-c_{\rm s}^{2}\bigg)^2.
\end{equation}

The bulk viscosity to entropy density ratio can be written as \cite{Sahoo:2017umy},
\begin{equation}
\label{eq12}
\zeta/s = 15\frac{\eta}{s}\bigg(\frac{1}{3}-c_{\rm s}^{2}\bigg)^2.
\end{equation}
We have estimated the values of $\zeta/s$ by using the above expression.

\subsection{Bulk Modulus}

Bulk modulus (B) of a substance is a measure of its degree of resistance to any external compression. Accordingly, it is defined
as the infinitesimal change (increase) in pressure with respect to the resulting change (decrease) in the volume of the system.
Following the derivations from standard text books, it can be expressed as \cite{halliday},
\begin{equation}
\label{eq15}
B = -V\frac{\partial P}{\partial V}
\end{equation}
From Ref.~\cite{Sahu:2020nbu}, the isothermal compressibility can be written as,
 \begin{equation}
\label{eq16}
\kappa_{\rm T} = \frac{1}{\frac{\langle m_{\rm T}\rangle dN_{ch}/dy}{2\tau_{\rm pro}S_{\rm N}}- \frac{5T^{3}e^{-\xi}\xi}{3L}}. 
\end{equation}
The bulk modulus is the inverse of isothermal compressibility,
 \begin{equation}
\label{eq17}
B = \frac{1}{\kappa_{\rm T}}.
\end{equation}
The bulk modulus gives information about the coupling of the constituents in a system -- higher is the value of bulk modulus, higher is the degree of coupling of the constituents in the system.

With these formulations of the observables in hand and their expressions in the framework of CSPM, we shall proceed to study them in the present context for various collision species and energies available at the LHC using the inputs from experimental data (transverse momentum spectra).

\section{Results and Discussion}
\label{res}

\begin{figure}[ht!]
\includegraphics[scale = 0.45]{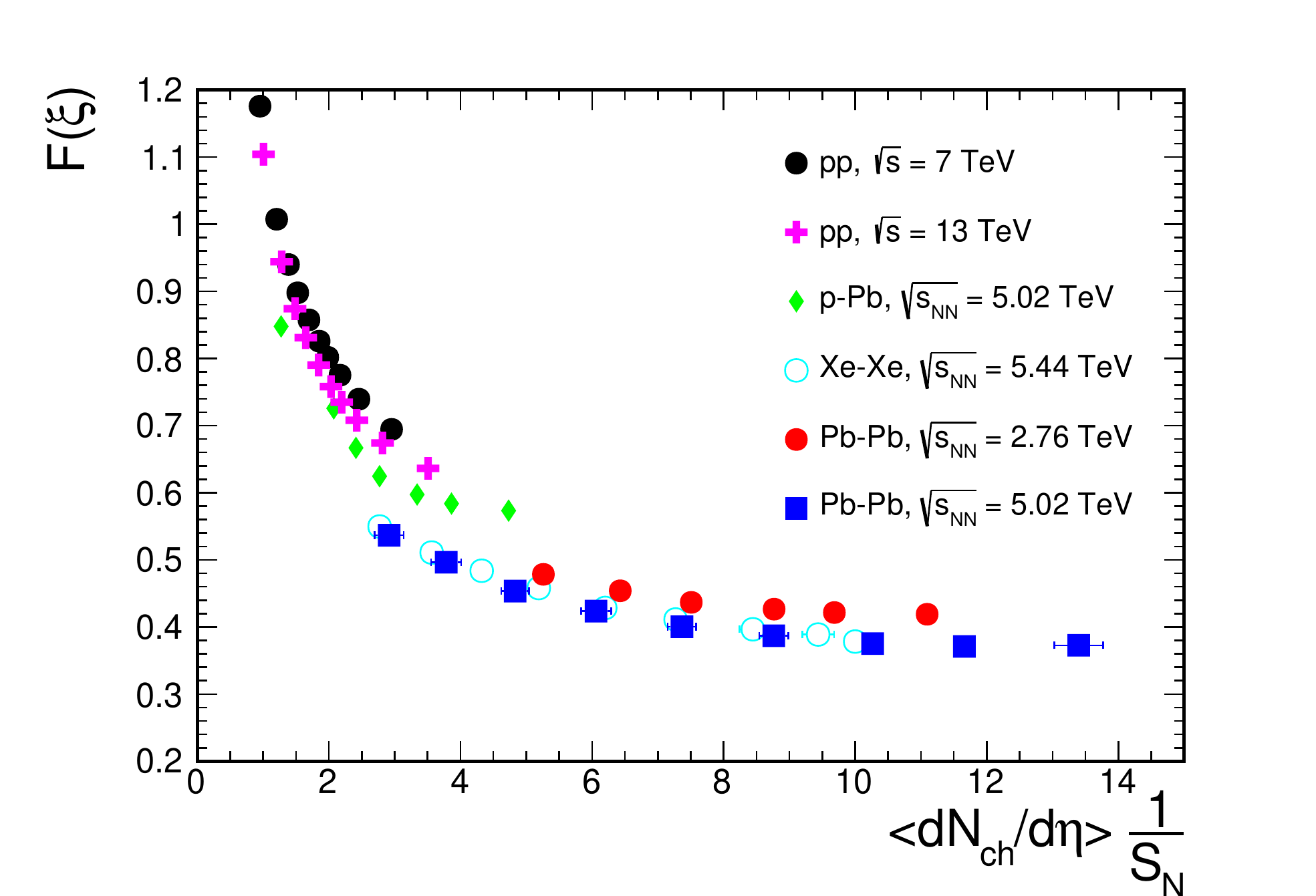}
\caption{(Color Online) Color suppression factor as a function of charged particle multiplicity density normalized by nuclear overlap area for $pp$ collisions at $\sqrt{s}$ = 7 and 13 TeV, $p$-Pb collisions at $\sqrt{s_{\rm NN}}$ = 5.02 TeV, Xe-Xe collisions at $\sqrt{s_{\rm NN}}$ = 5.44 TeV, Pb-Pb collisions at $\sqrt{s_{\rm NN}}$ = 2.76 and 5.02 TeV. The values for the $\langle dN_{\rm ch}/d\eta \rangle$ values are taken from Refs.~\cite{ALICE:2019etb,Acharya:2018orn,Abelev:2013vea,Acharya:2019yoi,Acharya:2019rys,Acharya:2018hhy,Acharya:2018egz}.}
\label{fig1}
\end{figure}

\begin{figure*}[ht!]
\centering
\includegraphics[scale = 0.44]{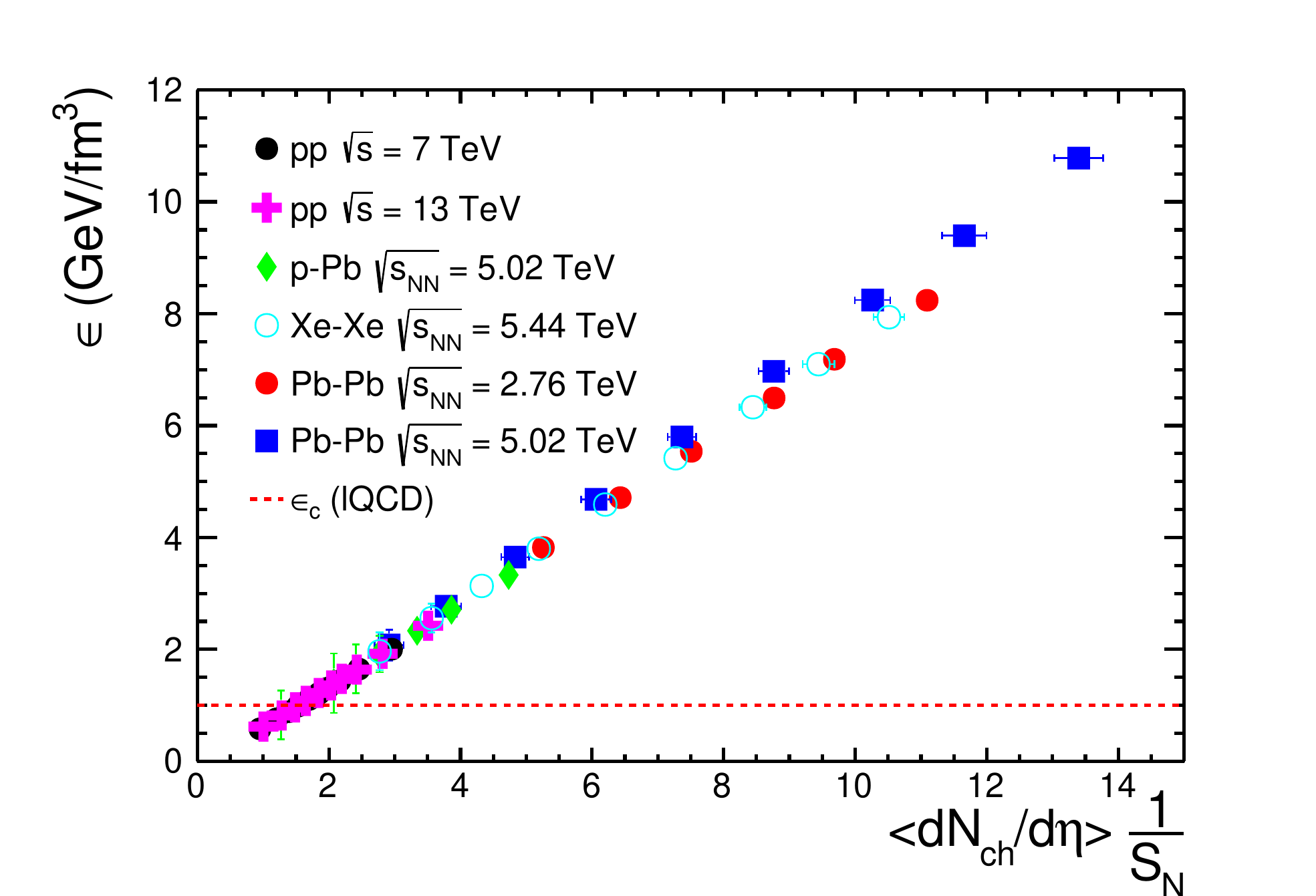}
\includegraphics[scale = 0.44]{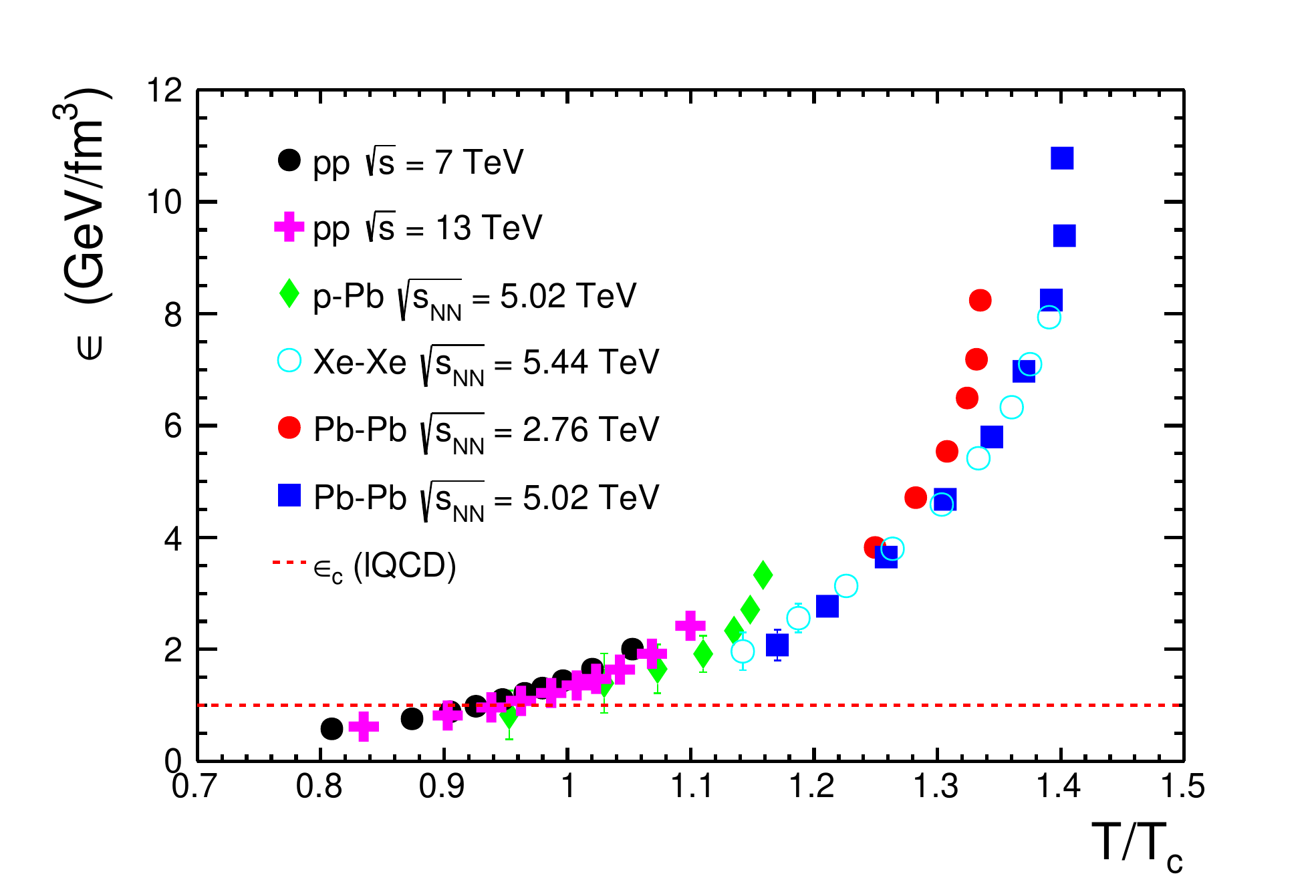}
\caption{(Color Online) Initial energy density as a function of charged particle multiplicity density normalized by nuclear overlap area (left panel) and as a function of initial percolation temperature normalized by critical temperature (right panel) for $pp$ collisions at $\sqrt{s}$ = 7 and 13 TeV, $p$-Pb collisions at $\sqrt{s_{\rm NN}}$ = 5.02 TeV, Xe-Xe collisions at $\sqrt{s_{\rm NN}}$ = 5.44 TeV, Pb-Pb collisions at $\sqrt{s_{\rm NN}}$ = 2.76 and 5.02 TeV. All the points are calculated from CSPM. The values for the $\langle dN_{\rm ch}/d\eta \rangle$ are taken from Refs.~\cite{ALICE:2019etb,Acharya:2018orn,Abelev:2013vea,Acharya:2019yoi,Acharya:2019rys,Acharya:2018hhy,Acharya:2018egz}.}
\label{fig8}
\end{figure*}

Figure \ref{fig1} shows the color suppression factor F($\xi$) as a function of charged particle multiplicity density normalized by nuclear overlap area. We observe that for low-multiplicity $pp$ collisions, F($\xi$) has the highest value and it goes on decreasing with the increase in the charged particle multiplicity density. The trend of the plot agrees with Refs.~\cite{Hirsch:2018pqm,Mishra:2019iyg}. Figure \ref{fig1} acts as a quality assurance plot for this study. Using the formulation discussed in the previous section, let us now move to the results for different thermodynamic observables as a function of charged particle multiplicity density normalized by nuclear overlap area. The final state charged particle density has been used as an event classifier in LHC $pp$ collisions 
\cite{nature} and has the advantage of using the final state information irrespective of the collision energy and system size. The final state
multiplicity in a system is proportional to the effective energy of the system contributing to the particle production.
Also, the thermodynamic and transport properties are studied as a function of initial percolation temperature normalized by the critical/hadronization temperature i.e. 167.7 $\pm$ 2.8 MeV~\cite{Becattini:2010sk,Karsch1}. Using the same set of experimental data, reported in one of our previous works \cite{Sahu:2020nbu}, we found that the initial percolation temperature is higher than the critical/hadronization temperature for event multiplicity, $\langle dN_{\rm ch}/d\eta \rangle \geq$ 20. This will make the subsequent discussion very interesting with special focus on high-multiplicity $pp$ collisions to explore for the possible formation of QGP.

Finally a parallelism is made between the dependence of
the discussed observables as a function of final state event multiplicity normalized to the nuclear overlap area and the initial percolation temperature.

\begin{figure*}[ht!]
\centering
\includegraphics[scale = 0.44]{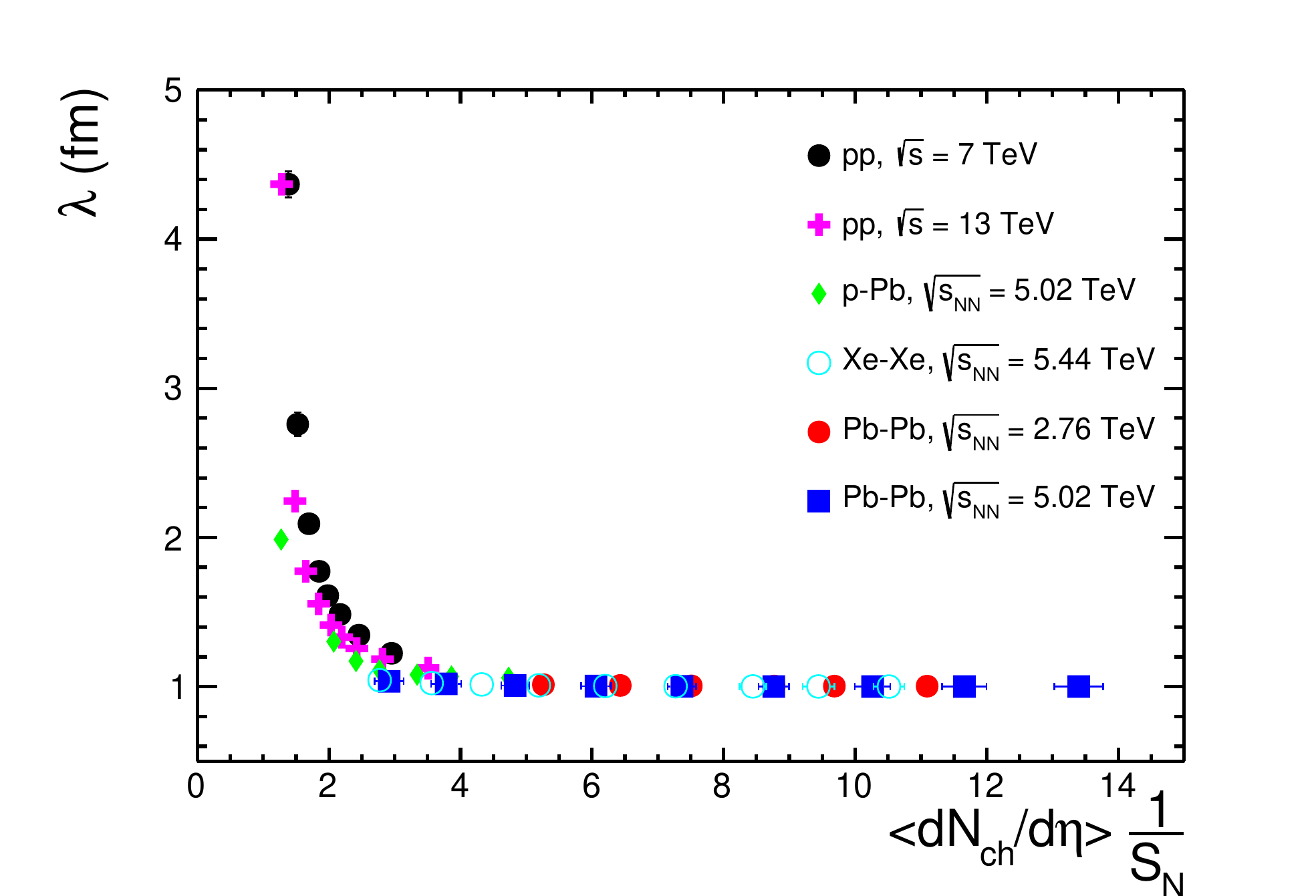}
\includegraphics[scale = 0.44]{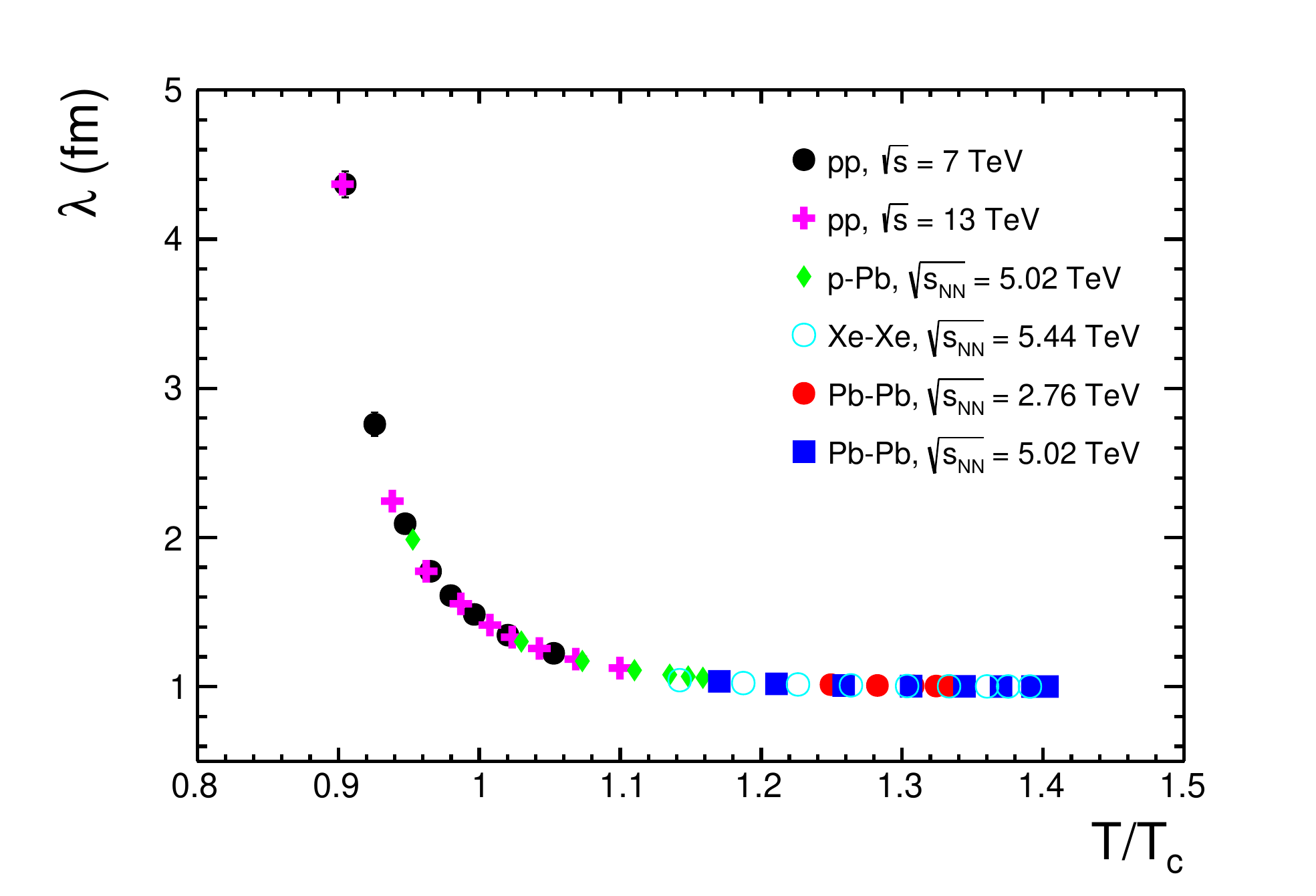}
\caption{(Color Online) Mean free path of the collision systems as a function of charged particle multiplicity density normalized by nuclear overlap area (left panel) and as a function of initial percolation temperature normalized by critical temperature (right panel) for $pp$ collisions at $\sqrt{s}$ = 7 and 13 TeV, $p$-Pb collisions at $\sqrt{s_{\rm NN}}$ = 5.02 TeV, Xe-Xe collisions at $\sqrt{s_{\rm NN}}$ = 5.44 TeV, Pb-Pb collisions at $\sqrt{s_{\rm NN}}$ = 2.76 and 5.02 TeV. All the points are calculated from CSPM. The values for the $\langle dN_{\rm ch}/d\eta \rangle$ are taken from Refs.~\cite{ALICE:2019etb,Acharya:2018orn,Abelev:2013vea,Acharya:2019yoi,Acharya:2019rys,Acharya:2018hhy,Acharya:2018egz}.}
\label{fig2}
\end{figure*}

\begin{figure*}[ht!]
\centering
\includegraphics[scale = 0.44]{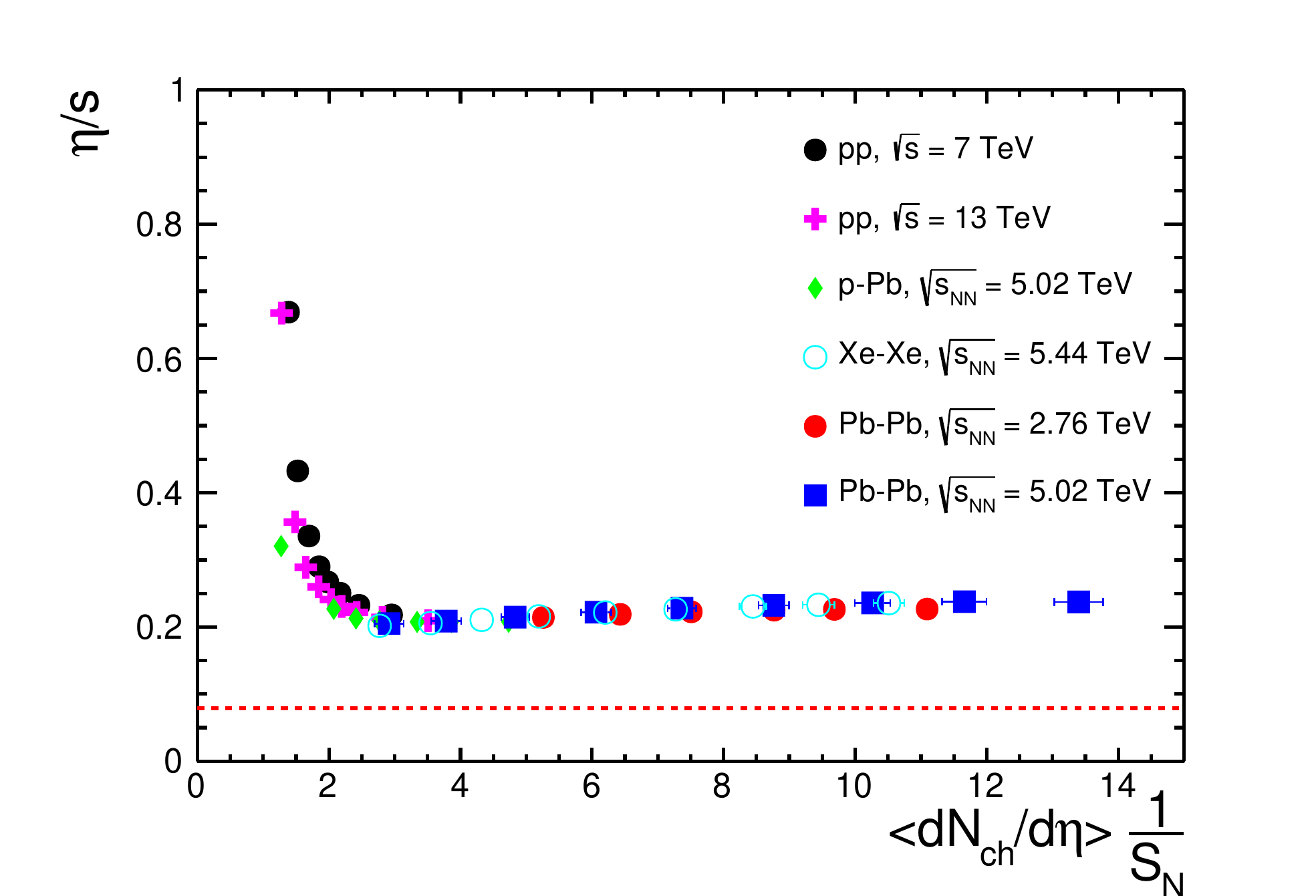}
\includegraphics[scale = 0.44]{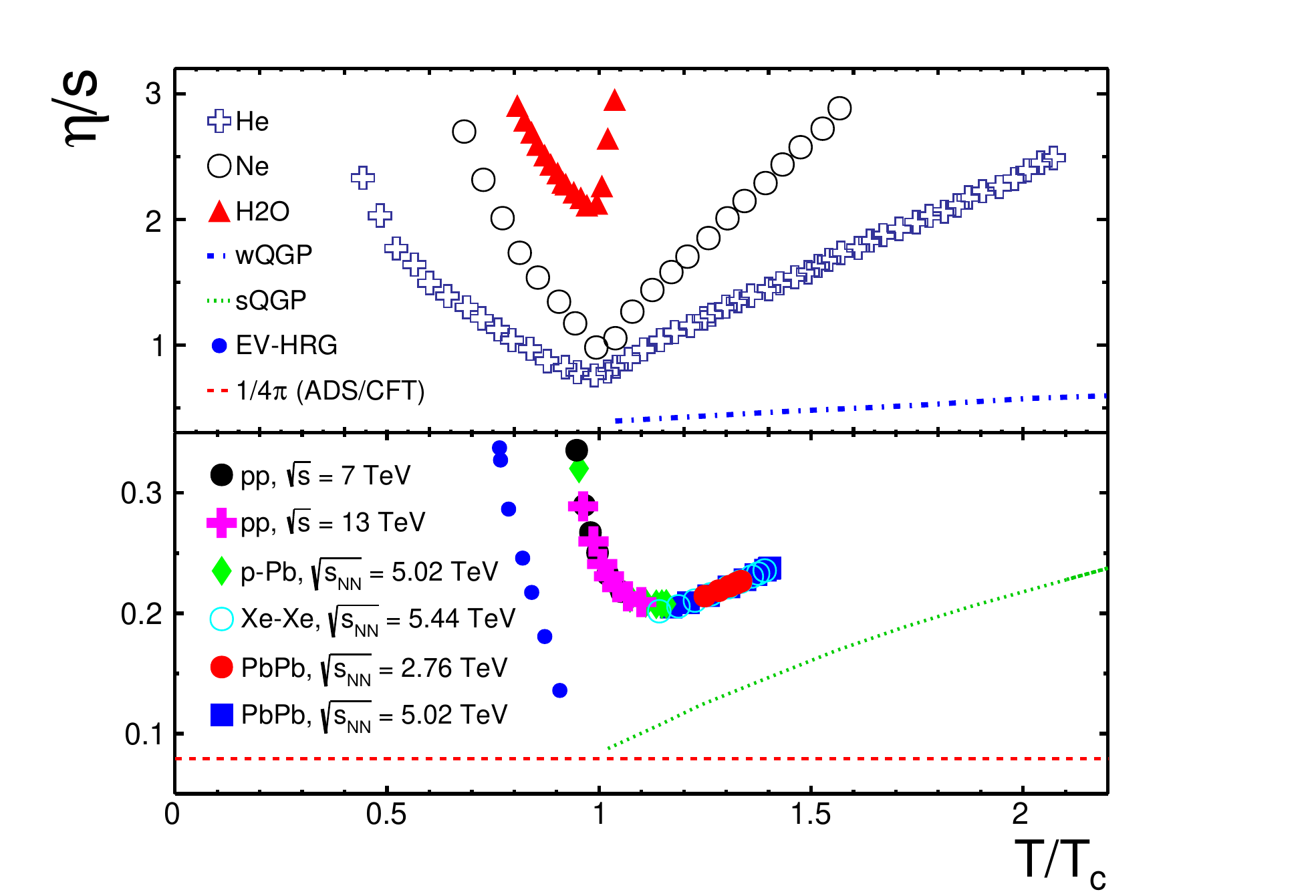}
\caption{(Color Online) The ratio of shear viscosity to entropy density as a function of charged particle multiplicity density normalized by nuclear overlap area (left panel) and as a function of initial percolation temperature normalized by critical temperature (right panel) for $pp$ collisions at $\sqrt{s}$ = 7 and 13 TeV, $p$-Pb collisions at $\sqrt{s_{\rm NN}}$ = 5.02 TeV, Xe-Xe collisions at $\sqrt{s_{\rm NN}}$ = 5.44 TeV, Pb-Pb collisions at $\sqrt{s_{\rm NN}}$ = 2.76 and 5.02 TeV are calculated from CSPM. The values for the $\langle dN_{\rm ch}/d\eta \rangle$ are taken from Refs.~\cite{ALICE:2019etb,Acharya:2018orn,Abelev:2013vea,Acharya:2019yoi,Acharya:2019rys,Acharya:2018hhy,Acharya:2018egz}. The results are compared with the predictions from AdS/CFT conjecture~\cite{Kovtun:2004de} and excluded volume hadron resonance gas model~\cite{Tiwari:2011km} as well as with physical systems like helium (He), neon (Ne) and water ($\rm H_2O$)~\cite{fluidcomp}. The results are also compared with the estimations for the weakly interacting (wQGP) and strongly interacting (sQGP) coupled QCD plasma~\cite{Hirano:2005wx}.}
\label{fig3}
\end{figure*}

Figure \ref{fig8} shows the initial energy density of the matter formed in high-energy collisions, calculated by using Eq.\ref{eq13}, as a function of charged particle multiplicity density normalized by nuclear overlap area (left panel) and initial percolation temperature normalized by critical temperature (right panel). Energy density is one among the most important observables, which gives insight into the bulk properties of hot QCD matter and characterizes the nature of the QCD phase transition. We observe that the initial energy density increases with the increase in charged particle multiplicity density.

When studied as a function of initial percolation temperature normalized by critical temperature in the right panel of Fig. \ref{fig8}, we observe an increasing trend with increase in temperature. In estimations of initial energy density, one always assumes the formation time of the quanta to be around 1 fermi/c, which can't be determined experimentally. This sets a lower bound
in the estimation of energy density, as one expects the formation time to be smaller with increase of collision energy. It is interesting to note that $pp$, $p$-Pb and heavy-ion collisions show similar energy densities at lower event multiplicity classes. However, towards higher multiplicities, one observes a sudden increasing trend in the initial energy density of the system. In view of this, one can infer that high-multiplicity $pp$ collisions at the LHC energies produce
initial energy densities, which are comparable with peripheral heavy-ion collisions and are well above the critical energy density of a deconfinement transition. This in fact sets the ball rolling to search for other evidence for a possible formation of QGP-droplets in these
high-multiplicity events.

\begin{figure*}[ht!]
\centering
\includegraphics[scale = 0.44]{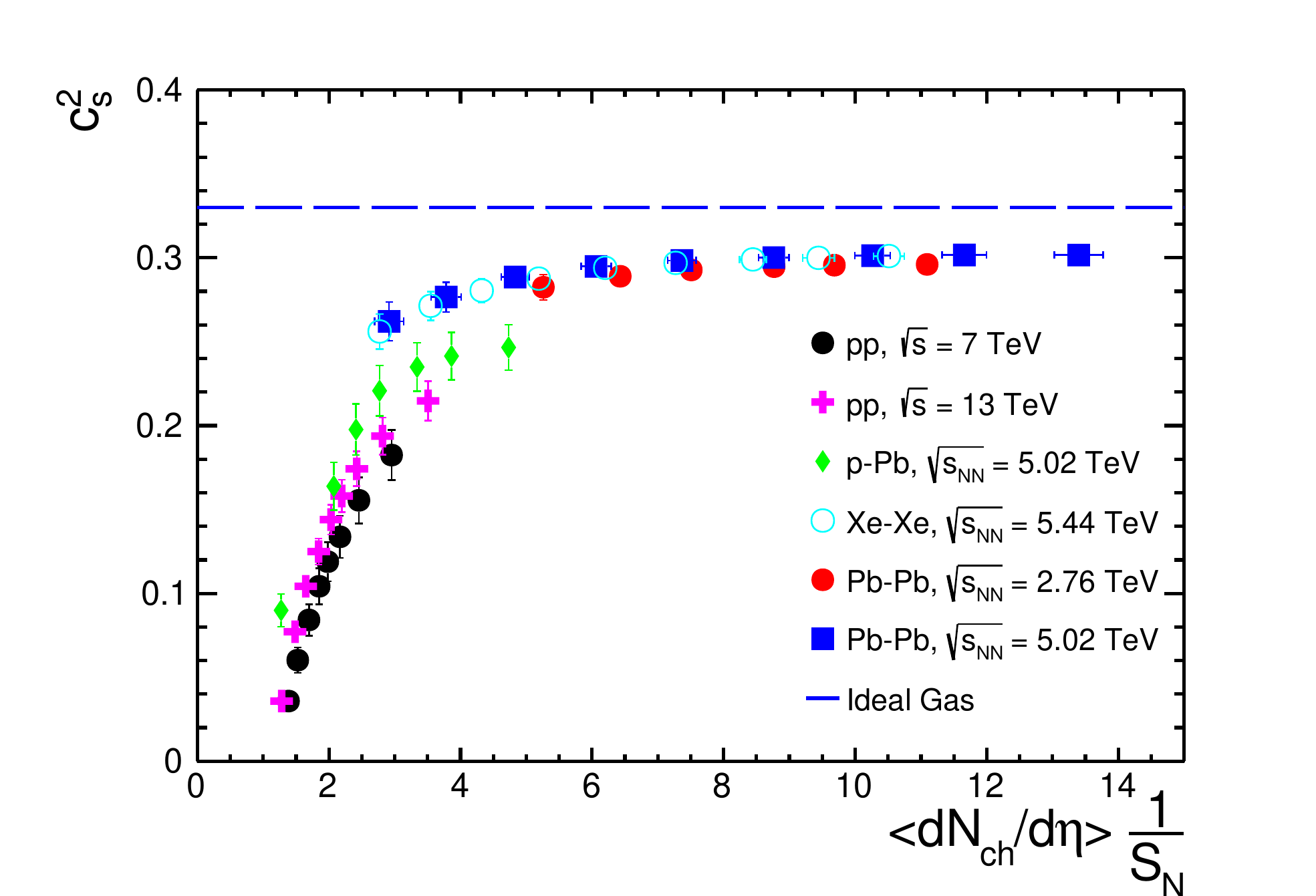}
\includegraphics[scale = 0.44]{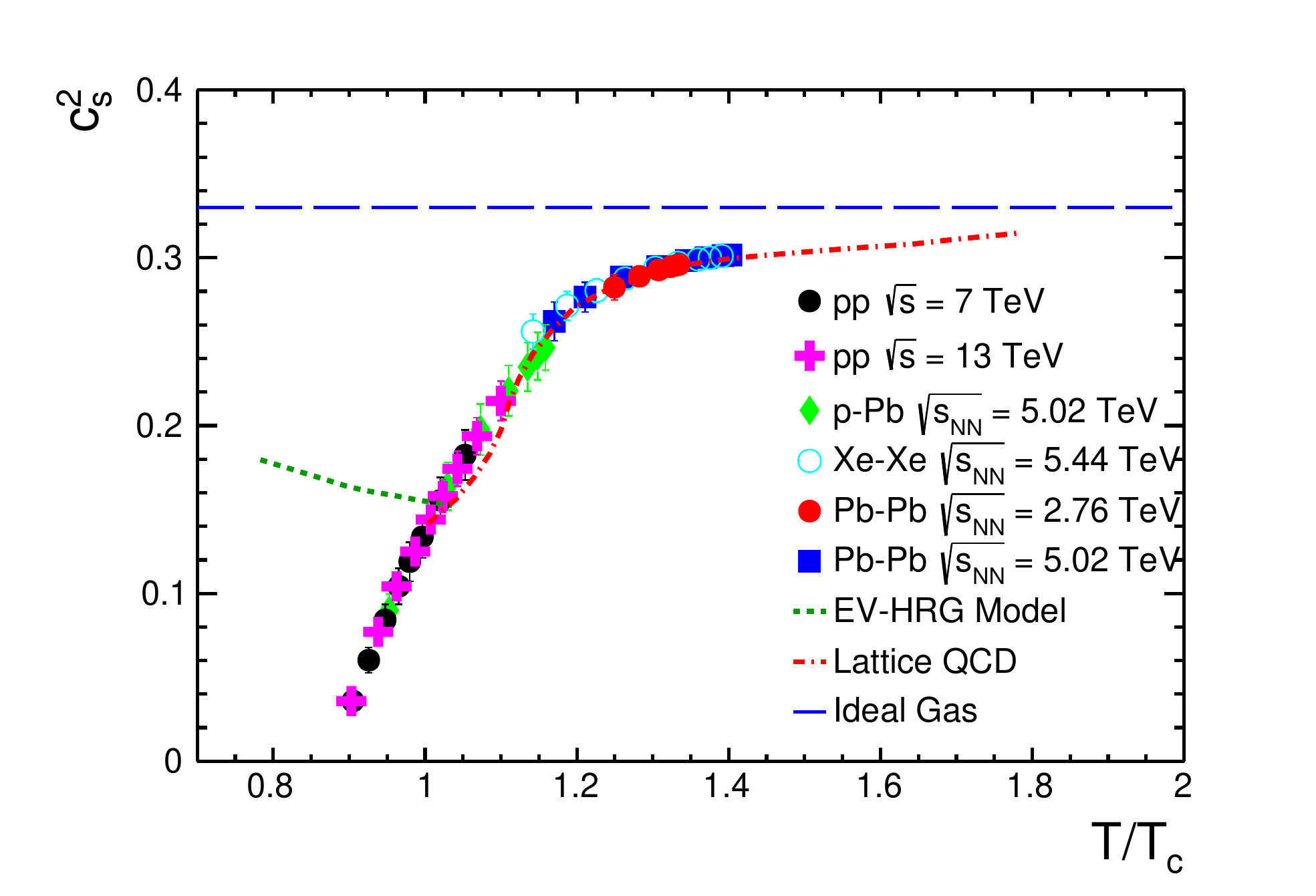}
\caption{(Color Online) Squared speed of sound as a function of charged particle multiplicity density normalized by nuclear overlap area (left panel) and as a function of initial percolation temperature normalized by critical temperature (right panel) for $pp$ collisions at $\sqrt{s}$ = 7 and 13 TeV, $p$-Pb collisions at $\sqrt{s_{\rm NN}}$ = 5.02 TeV, Xe-Xe collisions at $\sqrt{s_{\rm NN}}$ = 5.44 TeV, Pb-Pb collisions at $\sqrt{s_{\rm NN}}$ = 2.76 and 5.02 TeV. The values for the $\langle dN_{\rm ch}/d\eta \rangle$ are taken from Refs.~\cite{ALICE:2019etb,Acharya:2018orn,Abelev:2013vea,Acharya:2019yoi,Acharya:2019rys,Acharya:2018hhy,Acharya:2018egz}. The points are calculated from CSPM. The results are compared with the lattice QCD predictions~\cite{Bazavov:2009zn}, excluded volume hadron resonance gas model~\cite{Tiwari:2011km} and ideal gas limit.}
\label{fig5}
\end{figure*}

In Fig. \ref{fig2}, the mean free path ($\lambda$) of the system as a function of charged particle multiplicity density normalized by nuclear overlap area is shown and it is calculated using Eq.~\ref{eq9}. We see that for lower $\langle dN_{\rm ch}/d\eta \rangle \frac{1}{S_{\rm N}}$, $\lambda$ is higher and it goes on decreasing with the increase in $\langle dN_{\rm ch}/d\eta \rangle \frac{1}{S_{\rm N}}$. We know that the density of the system formed in high-energy collisions increases with the increase in charged particle multiplicity density. This is reflected in the $\langle dN_{\rm ch}/d\eta \rangle \frac{1}{S_{\rm N}} $ dependence of $\lambda$. At higher charged particle multiplicity density, we observe that $\lambda$ becomes the lowest,
saturating around one fermi. This is expected due to the fact that the exponential term in the expression of $\lambda$ in Eq.~\ref{eq9} becomes negligible with the increase of system size as $\xi$ increases with system size. This can be understood in the following way. Below critical percolation density ($\xi_{\rm c}$), with increase in temperature and charged particle multiplicity density, the string density increases rapidly, which would decrease the mean free path substantially. However, after the critical percolation density is reached, most of the area is already filled and thus the change in mean free path as a function of temperature and charged particle multiplicity density saturates. 

These observations go in line with our recent estimations of particle dependent mean free path estimated using the identified particle 
spectra in a Tsallis non-extensive framework \cite{Sahu:2020doe}.

\begin{figure*}[ht!]
\begin{center}
\includegraphics[scale = 0.44]{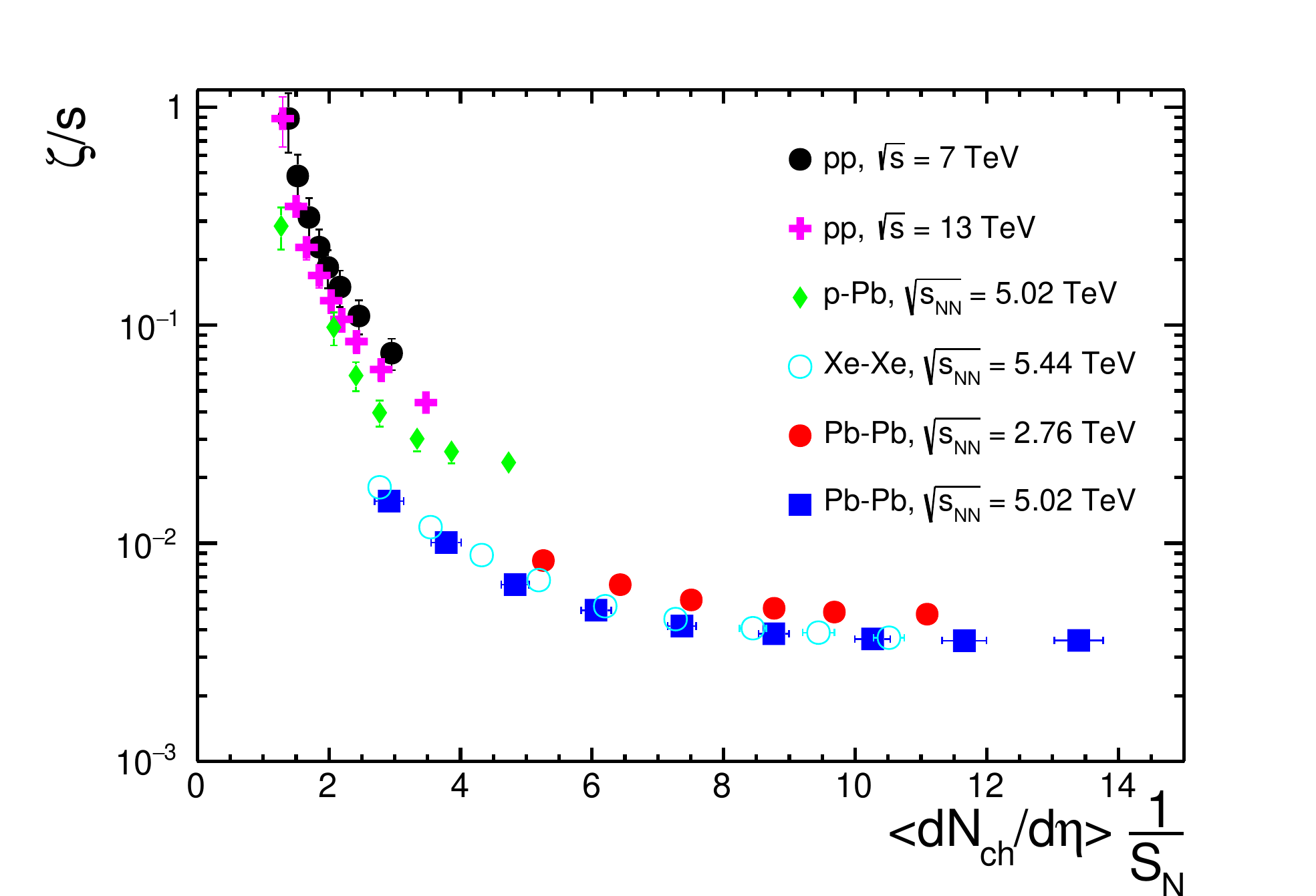}
\includegraphics[scale = 0.44]{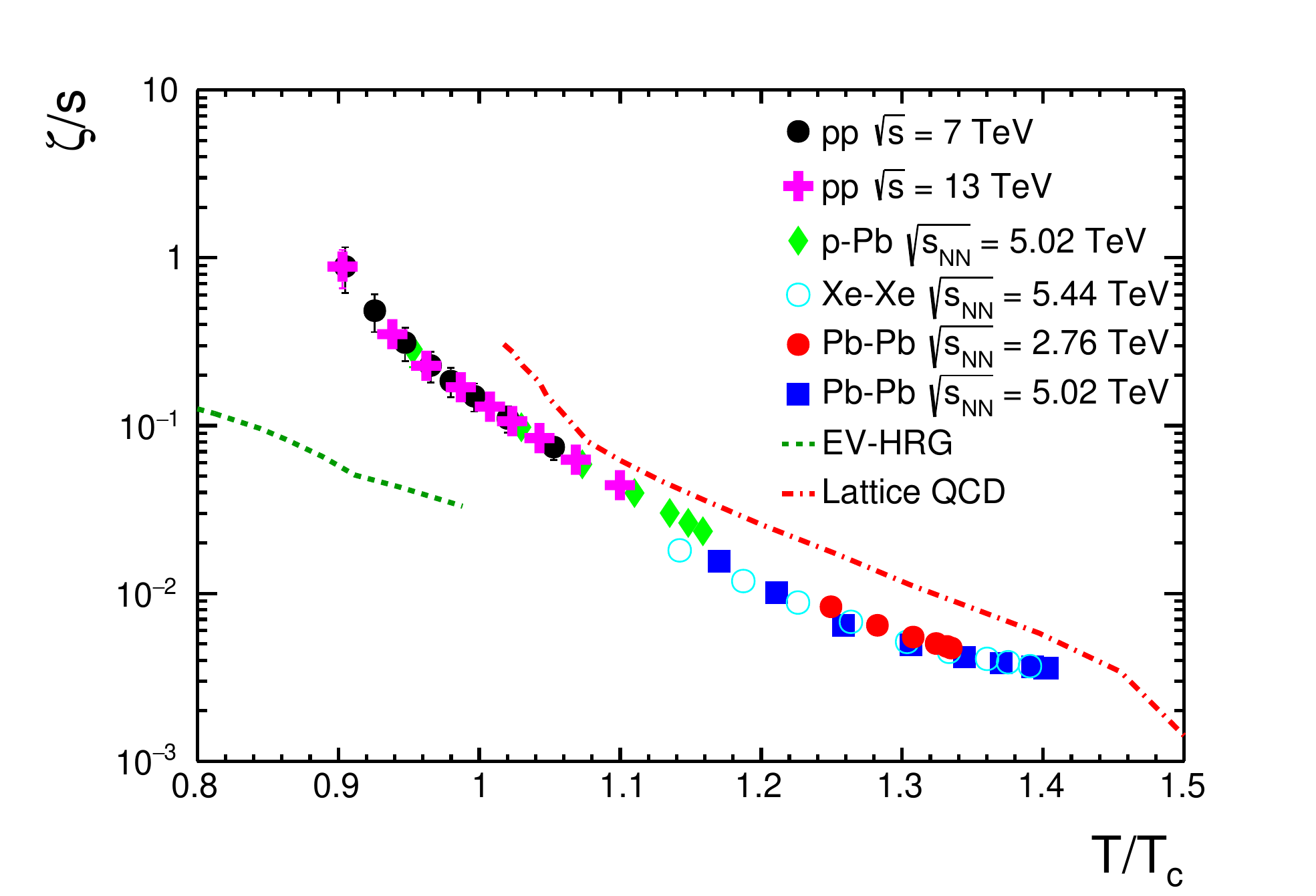}
\caption{(Color Online) The ratio of bulk viscosity to entropy density as a function of charged particle multiplicity density normalized by nuclear overlap area (left panel) and as a function of initial percolation temperature normalized by critical temperature (right panel) for $pp$ collisions at $\sqrt{s}$ = 7 and 13 TeV, $p$-Pb collisions at $\sqrt{s_{\rm NN}}$ = 5.02 TeV, Xe-Xe collisions at $\sqrt{s_{\rm NN}}$ = 5.44 TeV, Pb-Pb collisions at $\sqrt{s_{\rm NN}}$ = 2.76 and 5.02 TeV. The values for the $\langle dN_{\rm ch}/d\eta \rangle$ are taken from Refs.~\cite{ALICE:2019etb,Acharya:2018orn,Abelev:2013vea,Acharya:2019yoi,Acharya:2019rys,Acharya:2018hhy,Acharya:2018egz}. The points are calculated from CSPM. The results are compared with the lattice QCD predictions~\cite{Karsch:2007jc} and excluded volume hadron resonance gas model~\cite{Tiwari:2011km}.}
\label{fig6}
\end{center}
\end{figure*}

The left and right panels of Fig. \ref{fig3} show the shear viscosity to entropy density ratio calculated using Eq.~\ref{eq10} as a function of charged particle multiplicity density normalized by nuclear overlap area and initial percolation temperature, respectively. We observe that, $\eta/s$ starts from around 0.7 and slowly decreases with increase in charged particle multiplicity density. At $\langle dN_{\rm ch}/d\eta \rangle \frac{1}{S_{\rm N}}$ $\simeq$ 3, the $\eta/s$ becomes minimum and then starts increasing with the increase in the charged particle multiplicity density. The behavior of $\eta/s$ can be understood in the following way. Below critical percolation density ($\xi_{\rm c}$), with increase in temperature and/or charged particle multiplicity density, the string density increases rapidly, which in turn decreases the mean free path and $\eta/s$. Above the critical percolation density, two-third of the available transverse area is already filled. So, less area becomes available to be filled below critical percolation density. Thus, a small decrease in mean free path is compensated by the increase in temperature and/or charged particle multiplicity density, which results in a smooth increase of $\eta/s$. 
High-multiplicity $pp$ collisions and heavy-ion collisions showing similar values of $\eta/s$ suggest that the produced QGP is strongly coupled. There is a range of charged particle multiplicity density for which $\eta/s$ is minimum, approaching the KSS bound value ($\simeq 1/4\pi$). This is because the initial temperature of the system produced in this range of charged particle multiplicity density is almost the same. To verify this argument, the shear viscosity to entropy density ratio as a function of initial temperature is plotted in the right panel and we indeed observe the same. We can clearly see that at a certain temperature, the $\eta/s$ is minimum. This temperature has been identified as the critical or the hadronization temperature \cite{Lacey:2006bc}. After this, the trend again starts to increase which indicates the QGP phase. Our results are compared with the predictions of AdS/CFT conjecture~\cite{Kovtun:2004de} and excluded volume hadron resonance gas model~\cite{Tiwari:2011km} as well as with well-known fluids found in nature such as helium (He), neon (Ne) and water ($\rm H_2O$)~\cite{fluidcomp}. The results are also compared with the estimations for the weakly interacting (wQGP) and strongly interacting (sQGP) coupled QCD plasma~\cite{Hirano:2005wx}. It is observed that the minimum point of $\eta/s$ for the matter produced in high-energy collision is the smallest when compared to any other known fluid and close to the predictions from strongly interacting (sQGP) coupled QCD plasma. Also, the results are close to the minimum value of AdS/CFT conjecture. This makes the matter produced in high-multiplicity high-energy collisions to be the closest perfect fluid found in nature. Another significant conclusion can be drawn when the left and right panels of Fig. \ref{fig3} are compared with each other --  the change in $\eta/s$ is faster as a function of initial percolation temperature compared to the change as a function $\langle dN_{\rm ch}/d\eta \rangle$.

\begin{figure*}[ht!]
\includegraphics[scale = 0.44]{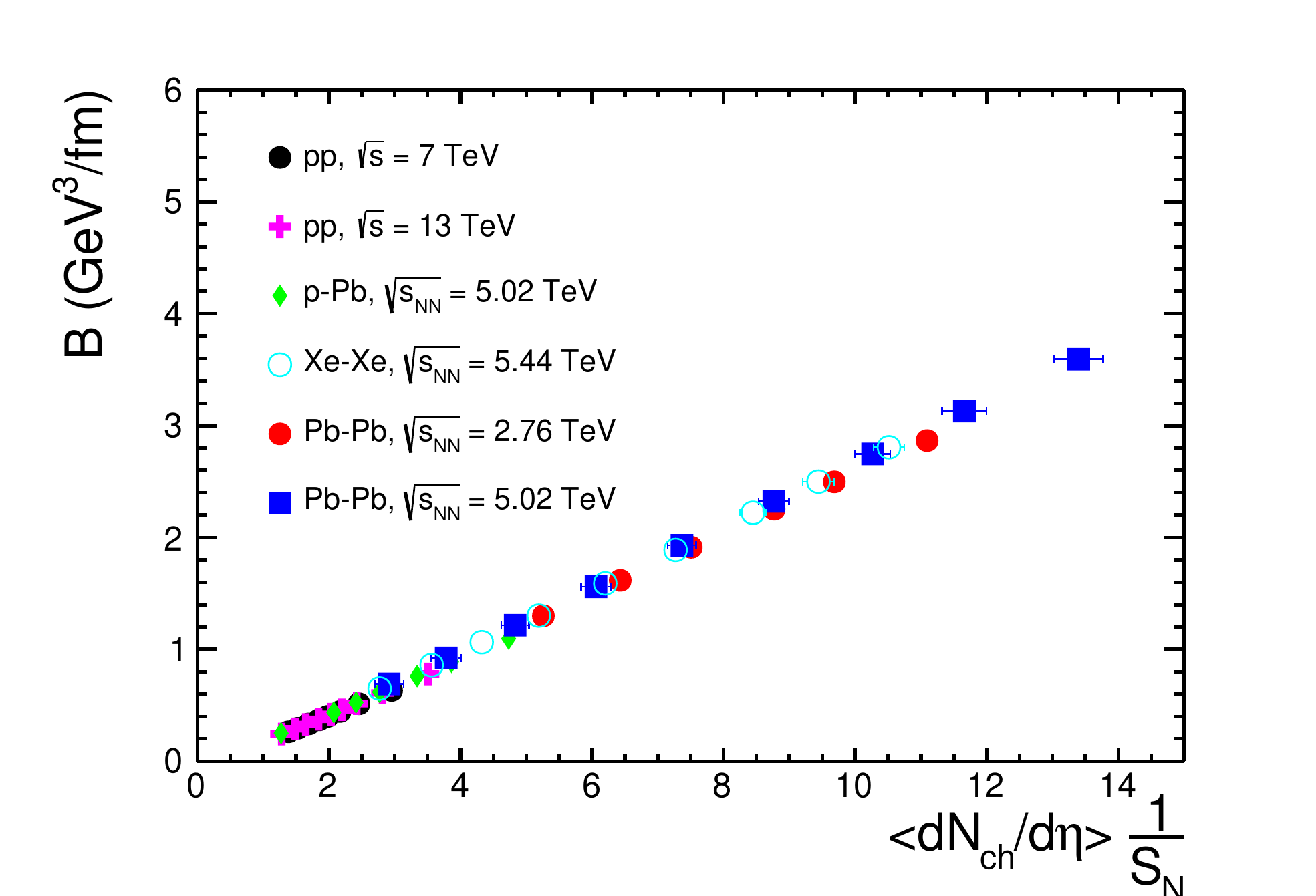}
\includegraphics[scale = 0.44]{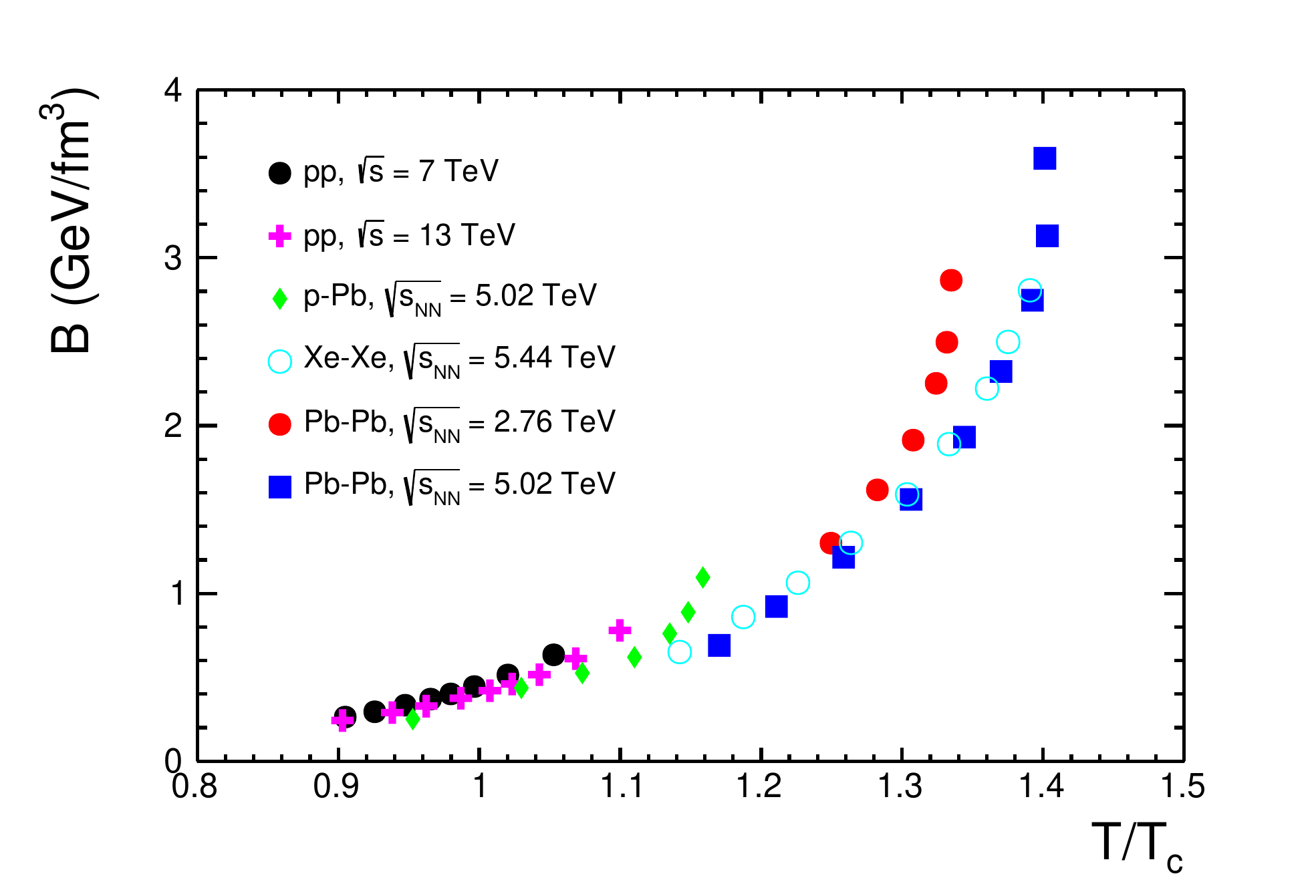}
\caption{(Color Online) Bulk modulus as a function of charged particle multiplicity density normalized by nuclear overlap area (left panel) and as a function of initial percolation temperature normalized by critical temperature (right panel) for $pp$ collisions at $\sqrt{s}$ = 7 and 13 TeV, $p$-Pb collisions at $\sqrt{s_{\rm NN}}$ = 5.02 TeV, Xe-Xe collisions at $\sqrt{s_{\rm NN}}$ = 5.44 TeV, Pb-Pb collisions at $\sqrt{s_{\rm NN}}$ = 2.76 and 5.02 TeV. The points are calculated from CSPM. The values for the $\langle dN_{\rm ch}/d\eta \rangle$ are taken from Refs.~\cite{ALICE:2019etb,Acharya:2018orn,Abelev:2013vea,Acharya:2019yoi,Acharya:2019rys,Acharya:2018hhy,Acharya:2018egz}.}
\label{fig9}
\end{figure*}

Figure \ref{fig5} shows the squared speed of sound, calculated by using Eq.\ref{eq11}, as a function of charged particle multiplicity density normalized by nuclear overlap area (left panel) and as a function of initial percolation temperature normalized by $T_{\rm c}$ (right panel). Here, we observe that for lower $\langle dN_{\rm ch}/d\eta \rangle \frac{1}{S_{\rm N}}$ the squared speed of sound is lower and it increases with the increase in the charged particle multiplicity density of the system. For most central heavy-ion collisions, the squared speed of sound approaches to the value of $c_{\rm s}^{2}$ of ideal massless gas, which is 1/3. This limit follows a massless blackbody equation of state as discussed in Landau hydrodynamic model of multiparticle production \cite{Landau}. Similar to the initial energy density variation
as a function of final state event multiplicity normalized to the nuclear overlap area, the squared speed of sound also shows a near-system-independent behavior, however, when the energy density shows a monotonic rise, the squared speed of sound saturates towards the ideal gas limit. 
The variation of $c_{\rm s}^{2}$ as a function of
normalized temperature is then compared to the results of lattice QCD predictions~\cite{Bazavov:2009zn} and excluded volume hadron resonance gas model~\cite{Tiwari:2011km}. The obtained results agree with the predictions from lattice QCD beyond $T = T_c$. 

Figure \ref{fig6} shows the variation of bulk viscosity to entropy density ratio, calculated by using Eq.\ref{eq12}, as a function of charged particle multiplicity density normalized by nuclear overlap area (left panel) and initial percolation temperature normalized by $T_{\rm c}$ (right panel). We observe that for lower charged particle multiplicities, $\zeta/s$ is higher and it goes on decreasing with the increase in $\langle dN_{\rm ch}/d\eta \rangle \frac{1}{S_{\rm N}}$. 
The predictions from lattice QCD~\cite{Karsch:2007jc} agrees to a good extent with the results from CSPM beyond $T = T_c$.

According to Newton-Laplace formula, the bulk modulus and density in a fluid, determine the speed of sound by the relation, $c_{\rm s} = \sqrt{\frac{B}{\rho}}$. We estimate the bulk modulus of the matter formed in high-energy collision by using the Eq. \ref{eq17}. Figure \ref{fig9} shows the bulk modulus as a function of charged particle multiplicity density normalized by nuclear overlap area (left panel) and initial percolation temperature normalized by critical temperature (right panel). We observe that at low $\langle dN_{\rm ch}/d\eta \rangle \frac{1}{S_{\rm N}}$, the bulk modulus is very small. However, as $\langle dN_{\rm ch}/d\eta \rangle \frac{1}{S_{\rm N}}$ increases, the bulk modulus also increases. This suggests that at low-multiplicity events, the produced matter exhibits a small change in pressure with the change in volume. However for higher multiplicity events, the produced matter exhibits larger change in pressure with the change in volume. This is because of the fact that high-multiplicity events show higher initial energy density, which controls the
effective pressure gradient and expansion of the system. An expected system size dependence as seen previously for initial
energy density and squared speed of sound is also seen for bulk modulus as a function of final state event multiplicity. 
 To have a better grasp on the concept of bulk modulus of the matter formed in high-energy collision, we compared it to the bulk modulus of known fluids such as water. The bulk modulus of water at room temperature is about $1.367 \times10^{-26}~\rm{GeV/fm^{3}}$ \cite{halliday}, which is 26 orders of magnitude smaller as compared to the bulk modulus of the matter formed in high-multiplicity events. For an ideal fluid the bulk modulus should be infinite. Although an actual ideal fluid doesn't exist, the matter produced in high-multiplicity collisions shows a bulk modulus of about 26 orders of magnitude higher than water, thus suggesting that it is the closest to an ideal fluid found in nature.
\\
\section{Summary}
\label{sum}
We study the thermodynamic and transport properties of the matter formed in $pp$, $p$-Pb, Xe-Xe and Pb-Pb collisions using the CSPM approach. These observables are studied as a function of final state charged particle multiplicity density in 
pseudorapidity, which is an event classifier used in LHC, normalized by nuclear overlap area and as a function of initial state percolation temperature normalized by critical temperature. The results from this work are compared to the results from well known models of QCD and also they are compared to the thermodynamic and transport properties of matter found in day-to-day life such as water. In summary, we observe 
that the matter formed in high-multiplicity events is nearest to a perfect fluid found in nature.  High-multiplicity $pp$ collisions and heavy-ion collisions showing similar values of $\eta/s$ is an
indication of a strongly coupled QGP formed in such collisions. 
For high-multiplicity $pp$ collisions, it is observed that the required critical initial energy density ($\epsilon_c \sim 1 ~ {\rm GeV/fm^3}$) and the critical deconfinement/hadronization temperature ($T_c \sim $ 167.7 $\pm$ 2.8 MeV) are achieved. The observations of a threshold in the final state charged particle multiplicity after which a change in the dynamics of the system can be observed are reported in various works~\cite{Sharma:2018jqf,Thakur:2017kpv,Sahu:2019tch,Sahu:2020swd}. These observations have profound implications in the search for QGP-droplets in high-multiplicity $pp$ collisions. The discussed observables, when studied as a function of charged particle multiplicity normalized by nuclear overlap area, show a minor level of non-universality in their trends for different collision systems. The reason for this non-universality is not clearly known.

The present study done across a range of energies and collision species at the
LHC, would shed light in characterizing the produced systems with special emphasis on final state charged particle
multiplicity density.

\section*{Acknowledgement} 
The authors acknowledge the financial supports  from  ALICE  Project  No. SR/MF/PS-01/2014-IITI(G) of Department of Science $\&$ Technology,  Government of India and DAE-BRNS Project No. 58/14/29/2019-BRNS of Government of India. We would like to thank Dr. Sushanta Tripathy for useful discussions and providing constructive 
suggestions during the preparation of the manuscript.

 \end{document}